\documentclass{llncs}
\usepackage{makeidx}
\usepackage{bibentry}
\usepackage{amsmath}
\usepackage{graphicx, subfigure}
\usepackage{algorithm}
\usepackage{algpseudocode}
\usepackage{algorithmicx}
\usepackage{tikz}
\nobibliography*
\begin{document}
\pagestyle{headings}

\mainmatter

\title{Evaluation and Improvement of Laruelle-Widgr\'{e}n Inverse Banzhaf Approximation}
\author{Frits de Nijs \and Daan Wilmer}
\institute{Delft University of Technology,\\
	\email{\{f.denijs,d.w.h.wilmer\}@student.tudelft.nl}
}

\maketitle

\begin{abstract}
The goal of this paper is to critically evaluate a heuristic algorithm for the Inverse Banzhaf Index problem by Laruelle and Widgr\'{e}n. Few qualitative results are known about the approximation quality of the heuristics for this problem. The intuition behind the operation of this approximation algorithm is analysed and evaluated. We found that the algorithm can not handle general inputs well, and often fails to improve inputs. It is also shown to diverge after only tens of iterations. We present three alternative extensions of the algorithm that do not alter the complexity but can result in up to a factor $6.5$ improvement in solution quality.
\end{abstract}

\section{Introduction}
Many decisions are made using voting. A good example of voting are the presidential elections of the United States of America. This follows a two-step process, which nicely illustrates two types of voting. In the first step the citizens vote in each state. Every vote has the same weight and the candidate with the most votes in that state wins that state. The second step illustrates another type of voting: weighted voting. In this step every state votes for the candidate that won in that state. However, it would not be fair if each state has the same vote: the state of California represents over 37 million citizens while little over 560,000 live in Wyoming. Therefore each state has a certain weight, represented by a number of electors. The new president is then chosen by the majority of electors.

With such weighted voting situations, and especially when they are used to elect one of the most powerful men on earth, it is the question how fair the voting is. We can measure this, for example, by using the Banzhaf power index \cite{banzhaf} and comparing that to a fair power distribution. Instead of trying to create a fair index -- which is much more a philosophical and political question, rather than an algorithmic problem -- we try to find a distribution of weights of which the power index matches a target power index.

For this we first introduce the problem in a formal way and present an overview of some existing algorithms for this problem. Then we will discuss and analyse one of the algorithms and try to improve it, after which we evaluate the algorithm and our modifications.

\section{Problem statement }

The first step to measuring power is modeling a weighted voting situation, using a weighted voting game (WVG). A weighted voting game consists of a set $N$ of $n$ players $p_1, p_2,\ldots p_n$, each with a voting weight $w_1,w_2,\ldots w_n$, along with a \emph{quota} $q$. We write a WVG as $[q; w_1,\ldots,w_n]$. A coalition $C$ is a subset of players, and every coalition has a value $v(C) \in \{0,1\}$. A coalition with value $1$ is called winning, and a coalition with value $0$ is called losing.

A generally accepted (though not the only) method to measure \emph{a priori} power is the Banzhaf power index \cite{banzhaf}. It measures the power of a player $i$ by dividing the number of coalitions of \emph{other} players for which player $i$ is \emph{critical} (meaning that the coalition is losing, and that player $i$ can make it winning by joining it), by the total number of coalitions of other players \cite{basepaper}. Or in mathematical terms:
\begin{equation*}
\tilde{\beta_i} = \frac{1}{2^n - 1} \sum_{C \subseteq N} (v(C) - v(C \backslash \{p_i\})).
\end{equation*}

Often not the regular Banzhaf index is used, but the normalized version \cite{basepaper}. This abstraction is made when it is not interesting in how many cases players can actually exert power, but only how the power is distributed among players. Because that is what we need, we use the normalized Banzhaf index as well:
\begin{equation*}
\beta_i = \frac{\tilde{\beta_i}}{\sum\limits_{j = 1}^n \tilde{\beta_j}}
\end{equation*}

Just computing the power of a weighted voting game is an NP-hard problem \cite{prasadKelly} (see \cite{keijzerSurvey} for a thorough survey of problems related to power indices and algorithms for solving them). However, our goal is not to measure power. Our goal is to find a quota and an assignment of weights, such that the power is distributed the way we want. Because we use the Banzhaf power index to define the power distribution, our problem is the inverse Banzhaf problem. Formally written it looks as follows:

\begin{quote}
Given a target power index $t=(t_1,t_2,\ldots,t_n)$, find a weighted voting game (in weighted representation) $g=[q;w_1,\ldots,w_n]$ such that the Banzhaf index of $g$ is as close to $t$ as possible, according to some distance measure.
\end{quote}

\section{Current algorithms }

De Keijzer, Klos and Zhang propose a method to enumerate all weighted voting games for a given number of agents \cite{enumeration}. This algorithm can be used to find a voting game that is closest to a given target vector $t$: enumerate over all possible games, calculate the Banzhaf index for every game and store every game for which the resulting Banzhaf index is closer to $t$ according to the distance measure than the best game found so far. Finally, output the best game found.
Their algorithm runs in $O(2^{n^2 + 2n})$, and calculating the Banzhaf index takes $O(2^n)$ time. The combined algorithm is therefore $O(2^{n^2 + 2n} \cdot 2^n)=O(2^{n^2 + 3n})$. For an exact enumeration approach this is a significant result, however it is still highly intractable. Therefore, we think the algorithm is not really practical for larger instances (for example for computing the weights for the $100$ shareholders of a company).

Fatima, Wooldridge and Jennings designed an iterative approximation algorithm \cite{fatima}. The algorithm shifts small amounts of power from players that have too much power, to players that have too little power, according to a comparison of the game's power index with the target. The authors approximate the power index  using randomisation \cite{shaheenapprox}. As a consequence the algorithm is $O(n^2)$ per iteration. Their update rules have a property that makes the algorithm \emph{anytime}: it can be stopped at any iteration and every iteration gives a better or equal result. Unfortunately, it is focused on the Shapley-Shubik power index \cite{shapleyshubik}---which is similar, but not equal to the Banzhaf power index.

Aziz, Paterson and Leech also designed an iterative approximation algorithm \cite{algorithm}. Their algorithm is used to approximate the Banzhaf index, but instead of calculating it directly they use generating functions \cite{bilbaofernandez}. This way the Banzhaf index can be calculated efficiently, but only if the weights are integer. They use interpolation of the current voting power and the desired voting power to determine the next set of weights, multiply them with a certain factor and then round them to integers. They don't provide an analysis of the approximation quality of their algorithm.

\subsection{Laruelle-Widgr\'{e}n }

In our paper we focus on the algorithm by Laruelle and Widgr\'en. It is a relatively simple algorithm that uses the fact that every input (the weight of the player) has a corresponding output (the power of the player) and that there is a correlation between input (weights) and output (power distribution) of the Banzhaf algorithm -- increasing or decreasing the input often leads to similar changes in the output.

Using that fact, Laruelle and Widgr\'en constructed an iterative algorithm. In every step, they update the weights of the players by calculating the Banzhaf index, calculating the ratio of the banzhaf index divided by the target power index per player and then dividing the weight of each player by its corresponding ratio. That way the weights are adjusted according to the error in the Banzhaf index.

The target vector is used as the first set of weights. This makes sense because the power and the weight distribution are roughly correlated. From this starting point it is possible to iterate as many times as desired. A distance threshold can be set so the algorithm will stop when it is close enough, and a maximum number of iterations can be given to make sure the algorithm takes only a limited amount of time.

A pseudocode version of this algorithm is given in Algorithm \ref{alg:base}. It takes as input the vector `target' (also called $t$ below) and the numbers `maxDistance' (the distance threshold) and `maxIterations' (the maximum number of iterations). The vectors weight, banzhafIndex and ratio are all vectors of equal size equal to the number of players.
\begin{algorithm}
\caption{Laruelle-Widgr\'en}
\label{alg:base}
\begin{algorithmic}[1]
\Require {The vector `target' is a normalized vector of size $n$, with $n > 0$.}
\State For each player $i$: $\text{weight}(i) \gets \text{target}(i)$
\State $\text{iterations} \gets 0$
\Repeat
	\State $\text{banzhafIndex} \gets \text{calculateBanzhaf(weight)}$
	\State For each player $i$: $\displaystyle \text{ratio}(i) \gets \frac{\text{banzhafIndex}(i)}{\text{target}(i)}$
	\State For each player $i$: $\displaystyle \text{weight}(i) \gets \frac{\text{weight}(i)}{\text{ratio}(i)}$
	\State $\text{distance} \gets \text{distance}(\text{banzhafIndex}, \text{target})$
	\State $\text{iterations} \gets \text{iterations} + 1$
\Until{$\text{distance} < \text{maxDistance} \lor \text{iterations} > \text{maxIterations}$}
\Ensure {weight is a vector of size n}
\end{algorithmic}
\end{algorithm}

The authors do not give any guarantees about the algorithm. In their paper it is shown to give a good approximation for some cases, but the general case is not analyzed.

\section{Analysis }

Besides the target power distribution $t$ (called $\beta_\text{fair}$ by the authors of the algorithm), the Laruelle-Widgr\'{e}n algorithm also requires a valuation function $v$. The target power distribution must be a vector inside the regular Simplex, i.e. a vector where the elements sum to $1$, because in the computation of the ratio, $t$ is compared to normalized Banzhaf vectors, which are all elements of the regular Simplex. The order of the elements in $t$ does not matter for the algorithm, as a reordering in the input simply leads to an equal reordering of the outputs. Thus, without loss of generality we may assume the elements in the vector to be in non-increasing order. Therefore, whenever we apply the algorithm to a vector $t$, we take $t$ to be an element of the \emph{ordered} regular Simplex.

The authors consider four different valuation functions in their evaluation of the EU voting system:
\begin{enumerate}
	\item Unanimity $v_u$, which returns a $1$ iff all players are in the coalition.
	\item Simple Majority $v_{\text{sm}}$, which returns a $1$ iff the size of the coalition exceeds $8$ in the EU case, or $\lceil \frac{n}{2} \rceil$ in general.
	\item Qualified Majority $v_{\text{qm}}$, which returns a $1$ iff the weight of the coalition exceeds $62$ out of $87$ points in the EU case, or quota $q$ in general.
	\item Qualified Majority with minimum size $v_{\text{qm+}}$, which returns a $1$ iff the weight of the coalition exceeds $62$ and the size exceeds $10$ in the EU case, or quota $q$ and size $s$ in general.
\end{enumerate}

Since the valuation functions $v_u$ and $v_{\text{sm}}$ always return the same Banzhaf power index regardless of the player weights, it is not interesting to consider them for the algorithm as no improvement can be made by manipulating the weights. Thus, for our analysis we will focus on the effect of $v_{\text{qm}}$ for different choices of $q$. The valuation function $v_{\text{qm+}}$ is also interesting to consider, however it is more general than $v_{\text{qm}}$ since $v_{\text{qm}}$ is equal to $v_{\text{qm+}}$ with $s=1$, which makes it harder to analyse. Section \ref{sec:expectq} presents our analysis of the expected effect of $q$ on the algorithm results.

We also want to investigate the effect of the initial weights determined in step $1$ of algorithm \ref{alg:base}. We will refer to this initial weight vector as $\omega_0$, the weight vector computed in iteration zero. The authors of the algorithm start with the initial weights $\omega_0$ set to the target power distribution $t$, without any further explanation as to why this choice was made. We propose an adaptation to the algorithm where the choice of $\omega_0$ is also passed as a parameter, and investigate the expected effect of three choices for $\omega_0$ in section \ref{sec:expectw}.

\subsection{Expected effect of $q$ \label{sec:expectq}}

The quota $q$ determines when a coalition has enough weight to win. There has been some work on the effect of manipulating the quota by Zuckerman et al.\ \cite{quota}. In particular, they show that the worst case difference in power under a change in $q$ is bounded for a given set of players. This is relevant when we consider what will happen to the Banzhaf power of the initial weight, as it means that small changes in $q$ for a fixed $\omega_0$ produce stable changes in its power.

Our intuition is that $q$ determines the average swinging coalition size, where we call a coalition swinging iff it is winning but contains at least one player which can make it losing by leaving the coalition.  A good example of this effect can be seen for the weights $\begin{bmatrix}\frac{1}{n}&\frac{1}{n}&\ldots&\frac{1}{n} \end{bmatrix}$. Under these weights the power vector is always the same since all players have equal weight. However the underlying game changes with $q$. When $q\le\frac{1}{n}$ every player can win on its own, and the average swinging coalition size is $1$ (when there is a coalition of two or more, no single player can make it losing of leaving because the remaining players are sufficient for winning, hence it is not a swinging coalition). When $\frac{1}{n}<q\le \frac{2}{n}$ every pair of players can win, and the average swinging coalition has size $2$, and so on.

If $q$ determines the average size of the swinging coalitions, then the best value of $q$ should be $0.5$. This is due to the fact that the number of possible coalitions of a given size $s$ out of $n$ behaves as a binomial distribution, with the largest number of possible coalitions occurring for $s=\frac{1}{2}n$. Having more coalitions to choose from is better because it gives a higher granularity in the Banzhaf vectors available, which should lead to a Banzhaf vector that is closer to $t$. Because of the nature of the binomial distribution, we expect $q$ to behave symmetrically.

\begin{figure}[ht]
	\begin{center}
	\subfigure[$q=\frac{1}{2}$]{
		\begin{tikzpicture}
			\fill [fill=red]    (0.4375,1.75) -- (-1.0780444566227678,-0.875) -- (0.4375,-0.875) -- cycle;
			\fill [fill=yellow] (0.4375,-0.875) -- (-1.0780444566227678,-0.875) -- (0.4375,-1.75) -- cycle;
			\draw [draw=green, thick] (0.4375,-0.875) -- (-1.0780444566227678,-0.875);
			\filldraw [fill=blue, draw=black] (-1.0780444566227678,-0.875)  circle [radius=.07] node [left]  {\scriptsize $\left[\frac{1}{2}\;\frac{1}{2}\;0\right]$};
			\begin{scope}[black, >=stealth, ->]
				\draw (-0.3,1.2) node[left] {\scriptsize $\left[1\;0\;0\right]$} -- (0.2,0.8); 
				\draw (0.6,-0.6) node[right] {\scriptsize $\left[\frac{3}{5}\;\frac{1}{5}\;\frac{1}{5}\right]$} -- (0.2,-0.875); 
				\draw (-0.3,-1.7) node[left] {\scriptsize $\left[\frac{1}{3}\;\frac{1}{3}\;\frac{1}{3}\right]$} -- (0.2,-1.2); 
			\end{scope}
		\end{tikzpicture}
	}
	\subfigure[$q=\frac{2}{3}$]{
		\begin{tikzpicture}
			\fill [fill=red]     (0.4375,1.75) -- (-0.5728629710818451,0.0) -- (0.4375,0.0) -- cycle;
			\fill [fill=green]   (-0.5728629710818451,0.0) -- (0.4375,0.0) -- (0.4375,-1.75) -- cycle;
			\fill [fill=blue]    (-0.5728629710818451,0.0) -- (-1.0780444566227678,-0.875) -- (0.4375,-1.75) -- cycle;
			\begin{scope}[black, >=stealth, ->]
				\draw (-0.3,1.2) node[left] {\scriptsize $\left[1\;0\;0\right]$} -- (0.2,0.8); 
				\draw (0.6,-0.6) node[right] {\scriptsize $\left[\frac{3}{5}\;\frac{1}{5}\;\frac{1}{5}\right]$} -- (0.2,-0.875); 
				\draw (-0.3,-1.7) node[left] {\scriptsize $\left[\frac{1}{2}\;\frac{1}{2}\;0\right]$} -- (-0.2,-1); 
			\end{scope}
			\filldraw [fill=yellow, draw=black] (0.4375,-1.75)              circle [radius=.07] node [right] {\tiny $\left[\frac{1}{3}\;\frac{1}{3}\;\frac{1}{3}\right]$};
		\end{tikzpicture}
	}
	\subfigure[$q=\frac{3}{4}$]{
		\begin{tikzpicture}
			\fill [fill=red]    (0.4375,1.75) -- (-0.3202722283113839,0.43750000000000044) -- (0.4375,0.43750000000000044) -- cycle;
			\fill [fill=green]  (-0.3202722283113839,0.43750000000000044) -- (0.4375,0.43750000000000044) -- (0.4375,-0.875) -- cycle;
			\fill [fill=yellow] (0.4375,-0.875) -- (0.058613885844308045,-1.5312499999999998) -- (0.4375,-1.75) -- cycle;
			\fill [fill=blue]   (-1.0780444566227678,-0.875) -- (-0.3202722283113839,0.43750000000000044) -- (0.4375,-0.875) -- (0.058613885844308045,-1.5312499999999998) -- cycle;
			\begin{scope}[black, >=stealth, ->]
				\draw (-0.3,1.2) node[left] {\scriptsize $\left[1\;0\;0\right]$} -- (0.2,0.8); 
				\draw (0.6,-0.2) node[right] {\scriptsize $\left[\frac{3}{5}\;\frac{1}{5}\;\frac{1}{5}\right]$} -- (0.3,-0.2); 
				\draw (-0.3,-1.7) node[left] {\scriptsize $\left[\frac{1}{2}\;\frac{1}{2}\;0\right]$} -- (-0.2,-1); 
				\draw (0.6,-1.7) node[right] {\scriptsize $\left[\frac{1}{3}\;\frac{1}{3}\;\frac{1}{3}\right]$} -- (0.3,-1.5); 
			\end{scope}
		\end{tikzpicture}
	}
	\end{center}
	\caption{Mapping of weights to power index in the ordered Simplex for $3$ players and varying $q$.}
	\label{fig:simplex2d}
\end{figure}
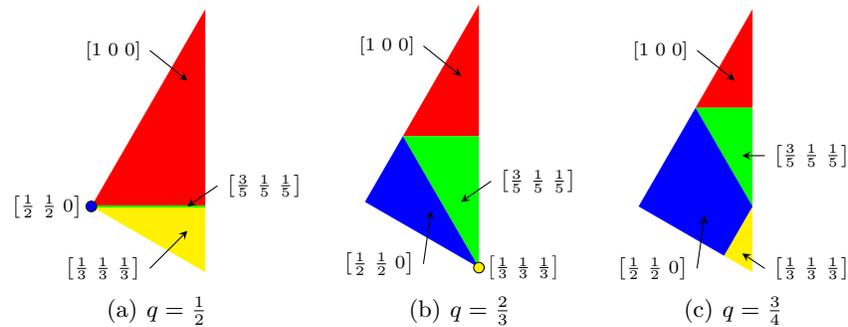

Another way to analyze the effect of $q$ is to look at its behavior for small numbers of players. In particular, when the number of players is three, we can visualize the Simplex in $2$D. This is done in figure \ref{fig:simplex2d}, for three different values of $q$. Because we're looking at the ordered Simplex, we only have to show the part of the regular simplex where the value (in these figures we show both weights and power) of the first player is not smaller than that of the second player, whose value is not smaller than the third player's. The figures show how different weight vectors (points in the ordered simplex) map to the four available Banzhaf vectors for 3 players (the four different colors), for three values of $q$ (Figures a through c). Only four Banzhaf vectors are attainable in WVGs on three players: $[1,0,0]$ (red, at the top), $[\frac{1}{2}, \frac{1}{2}, 0]$ (blue, on the left), $[\frac{3}{5}, \frac{1}{5}, \frac{1}{5}]$ (green, directly below red), and $[\frac{1}{3}, \frac{1}{3}, \frac{1}{3}]$ (yellow, directly below green).

As we can see in the figures, not all Banzhaf vectors are always represented by 2D planes or segments of the Simplex. When $q=0.5$ one Banzhaf vector maps to a line (a green line between the red and the yellow areas), while another (blue) maps to a point. When the first player has more weight than the quota of $0.5$, the others must have less and thus he becomes a dictator.
On the other hand, when he has less than $0.5$ the same holds for the other two and no-one can win without a second player. Thus only when the first player has weight exactly equal to $0.5$ can the other two vectors be reached. This is a challenge for any iterative algorithm, since they must potentially converge to a very small segment of the Simplex.

Also, from figure \ref{fig:simplex2d} we can see that the size of a segment of the Simplex that maps to a specific Banzhaf vector does not behave linearly for increasing $q$. Starting at $q=\frac{1}{2}$, the segment for $\beta=[\frac{1}{3}\:\frac{1}{3}\:\frac{1}{3}]$ first shrinks to a point towards $q=\frac{2}{3}$, and then grows to be a plane again when $q=\frac{3}{4}$.

The meaning of such a $2$D segment for the algorithm can be explained by looking at the ratio calculation in step six. For any Banzhaf vector and the given target power distribution, the ratio is uniquely defined. In other words, when the algorithm considers weights $\omega$, the update operation is uniquely defined by the Banzhaf vector that $\omega$ maps to. Where $\omega$ maps to is only controlled by parameter $q$.

The particular challenge for the algorithm is that the update step $\text{weight}=\frac{\text{weight}}{\text{ratio}}$ is not always defined. When the Banzhaf power vector contains a $0$ (red or blue in figure \ref{fig:simplex2d}) the ratio is also $0$, and the algorithm cannot continue. Such Banzhaf vectors appear only at the edges of the Simplex.

The effect of $q$ on the algorithm is thus not easily predicted. On the one hand, the algorithm does not always preserve weights, and therefore is not restricted to exploring just the weight vectors in the Simplex. On the other hand, the size differences between Banzhaf segments and existence of segments that stop the algorithm poses challenges.

\subsection{Expected effect of $\omega_0$ \label{sec:expectw}}

The authors of the original algorithm select the target vector $t$ as initial weights $\omega_0$ (the weights in the $0$th iteration) under the implicit assumption that this is a close guess to the final power distribution. However, we have seen in the previous section that this can lead to initial weights that produce a powerless player. This results in the algorithm dividing the weights by a zero ratio, and thus stopping after only one iteration. This may lead to poor approximation, and therefore this section attempts to find different guesses for $\omega_0$ to avoid this situation.

When considering choices for $\omega_0$ we may try the vertices of the ordered regular Simplex. However, all but one of the vertices contain at least one zero. The `outer' vertices of the regular Simplex thus immediately lead to a zero in the Banzhaf vector. For the same reason we should avoid the edges and higher dimensional `sides' of the regular Simplex. This leads us to the intuition that a good starting point is far from the edges.

The vertex of the ordered regular $n$-Simplex without a zero power player has the form $\begin{bmatrix}\frac{1}{n}&\frac{1}{n}&\ldots&\frac{1}{n} \end{bmatrix}$. This point is also the centroid of the regular Simplex. A centroid is a type of center, intuitively defined as the average of all vectors in the body, or alternatively as the center of its mass. The centroid of a general Simplex is computed as the normalized sum of its $n$ vertices ($v_0$ through $v_{n-1}$): \[ \frac{1}{n}\sum\limits_{i = 0}^{n-1} v_i \]

This point seems to be a good candidate for $\omega_0$, as it is the furthest from all sides. It is also the starting point used by the approximation algorithm of Fatima et al. \cite{fatima}. However, it actually leads to the same situation as choosing the target vector, only one iteration later. The explanation for this is that when all weights are equal the Banzhaf vector is the same as the weights. Then the computation for $\omega_1$, the weights in the first iteration, becomes (for each player):
\begin{equation*}
	\omega_1=\frac{\omega_0}{\text{ratio}}=\frac{\omega_0}{\frac{\beta_0}{t}}=\frac{\omega_0}{\frac{\omega_0}{t}}=t
\end{equation*}

However, we can still use the notion of centroid applied to the ordered Simplex. This point is also far from all sides, and furthermore appears to be in a region of the Simplex where there are many Banzhaf vectors. We base this intuition on the views of Kurz \cite{kurz} on the location of Banzhaf vectors in the ordered Simplex. Starting close to many Banzhaf vectors is desirable, because then each iteration is likely to jump to a new Banzhaf vector which in turn results in a slightly different ratio. Changing the ratio often introduces variance in the direction of update which intuitively leads to better convergence.

The interior vertex can also be used in another way: We may manipulate the starting weight vector based on the target vector $t$, so that it is less likely to start with a powerless player. Players are powerless when they only have a very small weight which in the Simplex means that some other players have a very large weight. We can smooth this initial weight distribution by averaging $t$ with the interior vertex, intuitively `pulling' the target $t$ towards the center, away from the zero edges. This initial weight vector $\omega_0$ is computed as
\[ \omega_0=\frac{t+\left[\frac{1}{n}\;\frac{1}{n}\;\ldots\;\frac{1}{n}\right]}{2}
\]

Because the relative effect of these alternative starting points on the performance of the algorithm is hard to assess analytically, we evaluate them empirically. We consider three choices of $\omega_0$ in our evaluation.
\begin{enumerate}
	\item Target, the starting point used by Laruelle and Widgr\'{e}n.
	\item Centroid, the center of mass of the ordered Simplex.
	\item Offset Target, the starting point halfway between the target and the centroid of the regular Simplex.
\end{enumerate}

Each choice attempts to satisfy different conditions. Starting point $1$ is likely closest to the desired power. Starting point $2$ is in an area that is dense in Banzhaf vectors. Starting point $3$ attempts to combine both aspects.

\subsection{Evaluation Metrics and Experimental Design}

In this section we present the metrics used to evaluate the effect of various parameter combinations. Since the purpose of the algorithm is to find a Banzhaf vector $\beta_{\text{opt}}$ that is closest to the input target vector $t$ it is natural to consider the distance between $t$ and the algorithm's output, for some notion of distance. For our evaluation we use the concept of Manhattan or Taxicab distance $d_1$, or $||.||_1$. We chose this distance norm because it relates to the results in \cite{kurz} on the lower bound on the distance between $t$ and $\beta_{\text{opt}}$, and because it is cheap to compute. Formally:
\[d_1(t,\beta)=\left|\left|t-\beta\right|\right|_1=\sum\limits_{i=0}^{n} \left|t_i - \beta_i\right| \]

The maximum value of $d_1$ in the regular Simplex is $2$, for example when comparing $\begin{bmatrix}1 & 0 & \dots & 0\end{bmatrix}$ and $\begin{bmatrix}0 & 1 & \dots & 0\end{bmatrix}$. In the ordered regular $n$-Simplex this maximum is $2 - \frac{2}{n}$, when comparing $\begin{bmatrix}1 & 0 & \dots & 0\end{bmatrix}$ and $\begin{bmatrix}\frac{1}{n} & \frac{1}{n} & \dots & \frac{1}{n}\end{bmatrix}$.

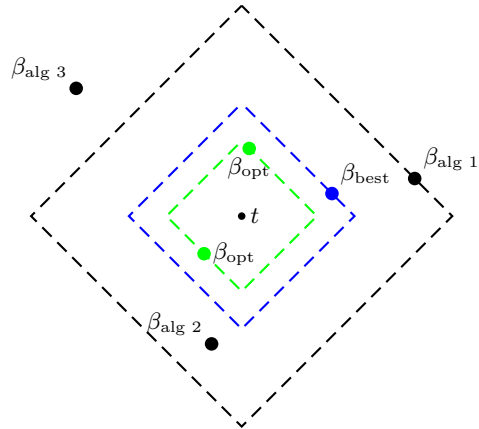
\begin{figure}
	\begin{center}
		\begin{tikzpicture}[dash pattern=on 5pt off 3pt]
			\draw [draw=green, dash phase=-1.8pt, thick] (0,1.0) -- (1.0,0) -- (0,-1.0) -- (-1.0,0) -- cycle;
			\draw [draw=blue,  dash phase=-2.2pt, thick] (0,1.5) -- (1.5,0) -- (0,-1.5) -- (-1.5,0) -- cycle;
			\draw [draw=black, dash phase=1pt, thick] (0,2.8) -- (2.8,0) -- (0,-2.8) -- (-2.8,0) -- cycle;
			\fill [fill=black] (0,0)      circle [radius=.05] node [right]       {$t$};

			\fill [fill=green] (.1,.9)    circle [radius=.09] node [below]       {$\beta_\text{opt}$};
			\fill [fill=green] (-.5,-.5)  circle [radius=.09] node [right]       {$\beta_\text{opt}$};

			\fill [fill=blue] (1.2,.3)    circle [radius=.09] node [above right] {$\beta_\text{best}$};

			\fill [fill=black] (2.3,.5)   circle [radius=.09] node [above right] {$\beta_{\text{alg }1}$};
			\fill [fill=black] (-.4,-1.7) circle [radius=.09] node [above left]  {$\beta_{\text{alg }2}$};
			\fill [fill=black] (-2.2,1.7) circle [radius=.09] node [above left]  {$\beta_{\text{alg }3}$};
		\end{tikzpicture}
	\end{center}
	\caption{Example of $d_1$ distance between target vector $t$ and related Banzhaf vectors in $2$D. All points on a diamond are equidistant to $t$, according to the $d_1$ distance metric.}
	\label{fig:metrics}
\end{figure}

In order to illustrate our evaluation metrics, Figure \ref{fig:metrics} presents an example of the distance between $t$ and a number of related Banzhaf vectors. In the figure, we consider the following Banzhaf vectors:
\begin{itemize}
	\item $\beta_{\text{opt}}$, an unknown optimal answer.
	\item $\beta_{\text{best}}$, a best known (closest) algorithm output found in a database of previously returned Banzhaf vectors.
	\item $\beta_{\text{alg }i}$, the algorithm output for certain parameter settings $i$.
\end{itemize}
There may be more than one optimal answer, for example when for two players the target $t=\begin{bmatrix}\frac{3}{4}&\frac{1}{4}\end{bmatrix}$ there exist two optimal Banzhaf vectors $\beta_{\text{opt}}=\begin{bmatrix}\frac{1}{2}&\frac{1}{2}\end{bmatrix}$ and $\beta_{\text{opt}}=\begin{bmatrix}1&0\end{bmatrix}$, both at distance $\frac{1}{2}$. Thus there can also be more than one best known Banzhaf vector, for example when both $\beta_{\text{opt}}$ are present in the database. A given algorithm parameter setting always produces the same result since the algorithm is deterministic.

We thus propose two metrics of interest:
\begin{enumerate}
	\item The \emph{relative improvement} obtained by using a parameter setting $2$ compared to $1$, \[ \frac{d_1(t,\beta_{\text{alg }1}) - d_1(t,\beta_{\text{alg }2})}{d_1(t,\beta_{\text{alg }1})}\]
	\item The \emph{error} in the output produced by a run of the algorithm defined as \[d_1(t,\beta_{\text{alg }i}) - d_1(t,\beta_{\text{opt}})\]
\end{enumerate}

The \emph{relative improvement} tells us something about the usefulness of a specific parameter setting. In other words, it tells us how the algorithm should be used to get the best possible results. In figure \ref{fig:metrics} we can see that the point for $\beta_{\text{alg }2}$ lies inside the black diamond indicating $d_1(t,\beta_{\text{alg }1})$. Thus, it is an improvement over $\beta_{\text{alg }1}$. In the computation for improvement we do not compare the points directly, but rather the distance to $t$, or the minimum Manhattan distance between $\beta_{\text{alg }2}$ and a point on the diamond for $\beta_{\text{alg }1}$ in figure \ref{fig:metrics}.

The improvement can be computed exactly since it only uses the known vectors. We note that the improvement may be negative, for example in figure \ref{fig:metrics} this is the case for $\beta_{\text{alg }3}$ compared to $\beta_{\text{alg }1}$. We consider an improvement to be significant if it exceeds $0.05$.

The \emph{error} tells us something about the general usefulness of the algorithm. The error is the minimum Manhattan distance between two points on the green and black diamonds in figure \ref{fig:metrics}. Since we do not know $\beta_{\text{opt}}$ the error must be estimated.

An upper bound on the error is $d_1(t,\beta_{\text{alg }i})$. This corresponds to the distance between a point on the black diamond and $t$ in figure \ref{fig:metrics}. This upper bound is tight, since it could be that $t = \beta_{\text{opt}}$, in which case $d_1(t,\beta_{\text{opt}})=0$. 

A lower bound on the error is $d_1(t,\beta_{\text{alg }i}) - d_1(t,\beta_{\text{best}})$. In figure \ref{fig:metrics} it is the minimum Manhattan distance between any two points on the black and blue diamonds. This lower bound is also tight, since we may have stored the optimal answer in which case $\beta_{\text{best}} = \beta_{\text{opt}}$.

The upper bound $d_1(t,\beta_{\text{alg }i})$ is a biased estimate of the actual error, since it is in general not the case that $t = \beta_{\text{opt}}$. This is a consequence of the fact that Banzhaf vectors are discrete. Kurz proves in \cite{kurz} that there exists a lower bound on the largest $d_1(t,\beta_{\text{opt}})$ of $\frac{1}{9}$, and conjectures that this bound is actually $\frac{14}{34}$.

The bias of the lower bound depends on the bias that originates from our method of obtaining Banzhaf vectors. The quality of the lower bound further depends on the percentage of all Banzhaf vectors we have in our database. We generated Banzhaf vectors by running the algorithm on random samples and storing the vector computed in each iteration of the algorithm (step four) until the database remained constant for $250$ samples. For $8$ players this resulted in $1,094,138$ Banzhaf vectors, which compared to the $2,730,164$ weighted voting games that exists for $8$ players \cite{keijzerArxiv,kurz} means that there exists at least $1$ vector for every $2.5$ games at this size.

The conjectured lower bound of $\frac{14}{34}$ on the largest distance between $t$ and $\beta_{\text{opt}}$ can be used to determine the size of a significant error for the upper bound estimate. We say that the upper bound error of an algorithm (parameter setting) is significant if the average value of the error exceeds $10\%$ of $\frac{14}{34}$, or $\frac{7}{185}$. Further, we say a change in the error is significant if the difference exceeds $1\%$ of the maximum value of $d_1$. Then a significant error for the lower bound estimate is at least $1\%$ of the maximum value of $d_1$.

\subsection{Experiments}
For our experiments we need a number of target vectors $t$ to apply the algorithm on. Since a target vector is a vector in the simplex, we produce samples $t$ by drawing a vector uniformly at random from the $n$-dimensional ordered regular Simplex. If function $U(0,1)$ returns a value drawn uniformly at random from the interval $[0,1]$, such a sample is constructed as:
\[
	\text{for}\:i \in (1, 2, ..., n),\:\tilde{t}(i) = -\log{\left(U(0,1)\right)}
\]
The elements from the vector $\tilde{t}$ are then normalized so that the resulting vector $t$ sums to $1$:
\[
	\text{for}\:i \in (1, 2, ..., n),\:t(i) = \frac{\tilde{t}(i)}{\sum\limits_{j = 1}^n \tilde{t}(j)}
\]
This results in a vector $t$ that is a random vector in the regular Simplex. To obtain a random vector in the ordered regular Simplex, the elements of the vector $t$ are sorted so that they are in descending order.

In order to evaluate empirically what choices of $q$ and $\omega_0$ produce the best results, we performed a number of experiments. For our experiments we drew $10,000$ samples from the ordered $8$D Simplex. On each sample we applied the algorithm for all choices of initial weight vector discussed in Section~\ref{sec:expectw}, and for $q$ ranging from $0.05$ to $0.95$ in steps of $0.05$. Each parameter combination was run for $50$ iterations.

\begin{figure}
	\centering
	\subfigure[$\omega_0$ set to target]{
		\includegraphics[width=36mm]{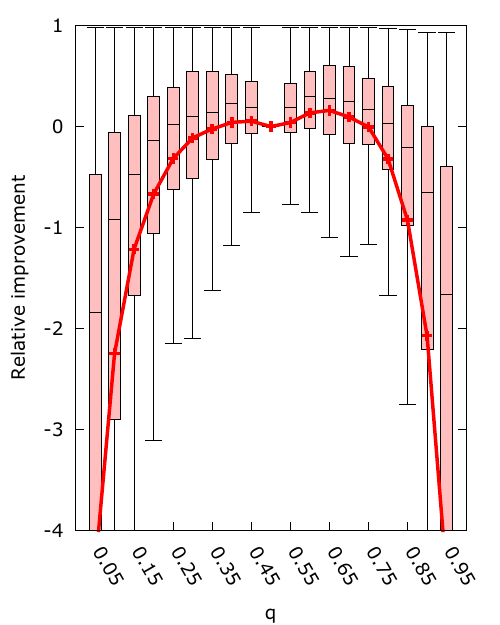}
	}
	\subfigure[$\omega_0$ set to centroid]{
		\includegraphics[width=36mm]{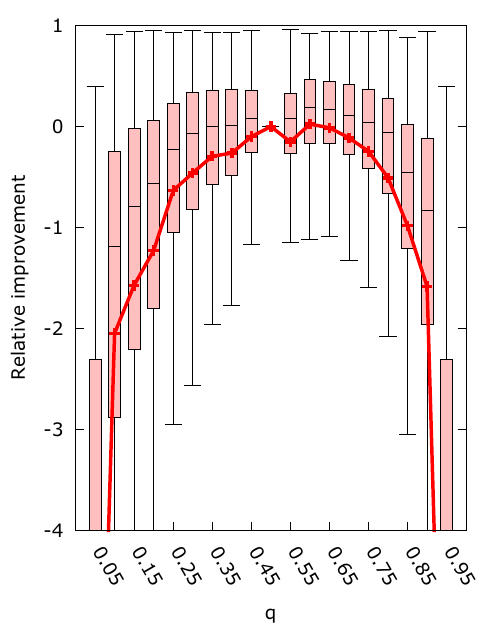}
	}
	\subfigure[$\omega_0$ set to offset]{
		\includegraphics[width=36mm]{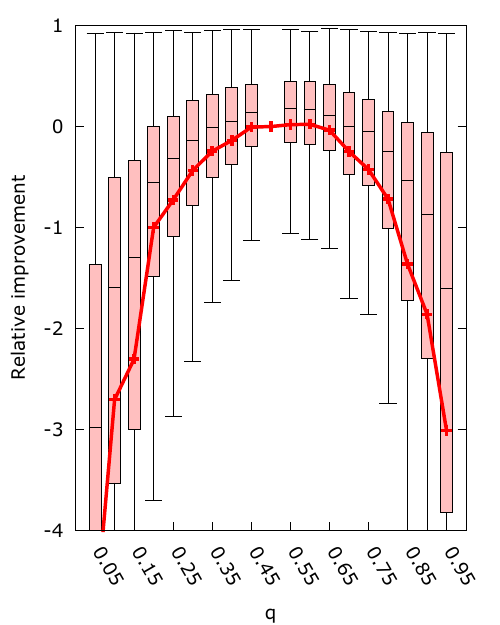}
	}
	\caption{Distribution of relative improvement versus the algorithm with $q=0.5$ for varying $q$. Mean relative improvement indicated with a line.}
	\label{fig:qparam}
\end{figure}

Figures \ref{fig:qparam} present the relative improvement obtained by running the algorithm with varying values of $q$ compared to running the algorithm with $q=0.5$ for each $\omega_0$. We expect $q=0.5$ to be the best setting, and that the effect of $q$ is symmetric. Thus we expect the relative improvement to be negative when $q \neq 0.5$, with equal magnitude on either side.

The figures show that the effect of $q$ on the mean improvement is generally symmetric. However, the algorithm does seem to perform better on values of $q > 0.5$, with the best mean performance occurring for $q = 0.65$ for $\omega_0$ set to the target, and $q=0.6$ for the other choices of $\omega_0$. This average improvement on $q > 0.5$ is only significant for $\omega_0$ set to the target. The asymmetry in the results may be caused by the update rule of the algorithm. In step six, the weight is updated as $\text{weight}=\frac{\text{weight}}{\text{ratio}}$. The ratio is independent of $q$ as it is computed from the normalized Banzhaf vector and the desired power. However the magnitude of the weights is dependent on $q$ since as $q$ becomes smaller an equal change in weights has a bigger relative effect.

Furthermore, choosing $q \neq 0.5$ results in significant improvements for the majority of samples when $q$ is chosen to be within $0.1$ of $0.5$ for all $\omega_0$. We expect that this is caused by the reduction in the area of the ordered Simplex that maps to the Banzhaf vector $[1\,0\,\ldots\,0]$. When $q=0.5$ a weight vector maps to $[1\,0\,\ldots\,0]$ whenever the first player has weight of $0.5$ or higher. As $q$ increases this area shrinks because the first player must have a higher weight to surpass $q$. Conversely, when $q$ decreases this area also shrinks because the other players need less weight to surpass $q$ themselves. We suspect this gives the algorithm more room to adjust the weights.

\begin{figure}
	\centering
	\subfigure[$\omega_0$ set to target]{
		\includegraphics[width=36mm]{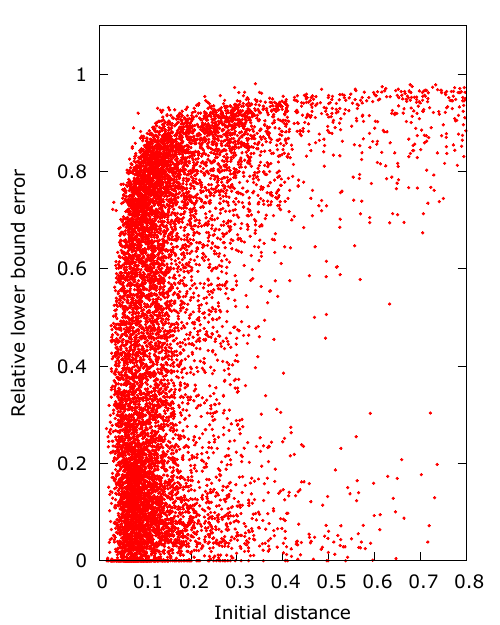}
	}
	\subfigure[$\omega_0$ set to centroid]{
		\includegraphics[width=36mm]{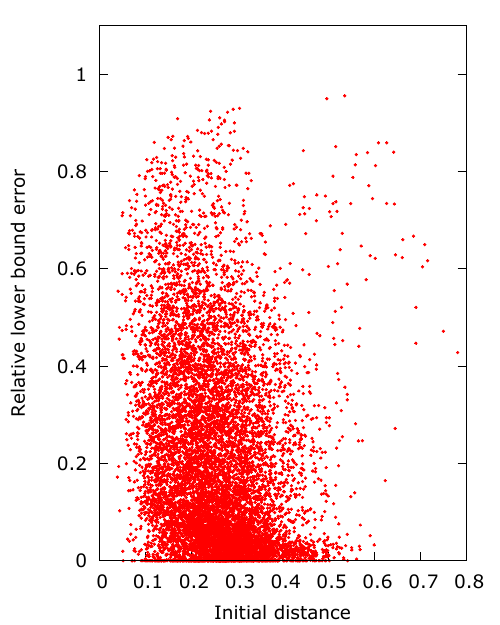}
	}
	\subfigure[$\omega_0$ set to offset]{
		\includegraphics[width=36mm]{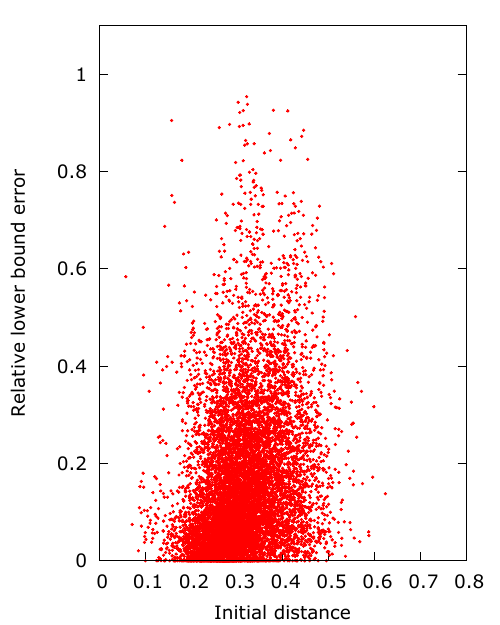}
	}
	\caption{Relative lower bound error compared to the initial distance for the varying $\omega_0$.}
	\label{fig:wparam}
\end{figure}

Figure \ref{fig:wparam} presents the effect of $\omega_0$ when $q=0.6$. The y-axis shows the relative lower bound error, computed as the lower bound on the error divided by the initial distance $d_1(t,\beta(\omega_0))$, and the initial distance $d_1(t,\beta(\omega_0))$ on the x-axis. This should show if there is a correlation between the error and the initial distance as produced by the choice of $\omega_0$.

Table \ref{tab:werror} presents the key features of the data in figure \ref{fig:wparam}. The first column shows the method for selecting the initial weight vector for the algorithm. The second column shows the average initial distance to $t$, and therefore the upper bound on the average initial error. The third column presents the worst case distance to $t$ after applying the algorithm, and thus the worst case error. Columns four and five present the average upper and lower bound error, where we note the average upper error is the average distance to $t$ after applying the algorithm.

\begin{table}
	\centering
	\begin{tabular}{ l | r r r r r }
		$\omega_0$ & \hspace{2mm}Initial & \hspace{2mm}Worst Error & \hspace{2mm}Error Upper & \hspace{2mm}Error Lower \\
		\hline
		Target & $0.1603$ & $0.7988$ & $0.1179$ & $0.0899$ \\
		Centroid & $0.2535$ & $0.6047$ & $0.0803$ & $0.0522$ \\
		Offset & $0.3146$ & $0.4228$ & $0.0833$ & $0.0552$ \\
	\end{tabular}
	\vspace{3mm}
	\caption{Effect of the different choices for $\omega_0$ on the error.}
	\label{tab:werror}
\end{table}

The figures show how the initial distance is affected by the choice of $\omega_0$, with $\omega_0$ set to the target having the smallest initial distance and offset having the largest. For $\omega_0$ set to either target or offset a higher initial distance generally results in a higher relative error, while for $\omega_0$ set to centroid the opposite appears to happen where the higher initial distances are improved more.

Another aspect of $\omega_0$ set to the target is that a large number of samples seem to be unimprovable, which is visible as a large number of samples with relative error near $1$. This may be caused by samples starting in a segment with a $0$ power player, which therefore cannot be improved beyond their initial guess.

The consequence of these unimprovable samples is that $\omega_0$ set as offset or centroid gives a significant improvement in both the upper and lower bound error compared to $\omega_0$ as target. However, the total magnitude of the error is still significant in all cases. The difference between offset and centroid is not significant, but offset does produce the lowest worst case distance to $t$, and it can thus be seen as the most robust. Overall we can conclude that starting close to the target is less important than starting in a position where the algorithm can improve the result.

\begin{figure}
	\centering
	\subfigure[Percentage of samples that stopped on a Banzhaf vector with at least one zero power player for different choices of $\omega_0$ and increasing number of iterations. Number of performed iterations on a log scale.]{
		\includegraphics[width=55mm]{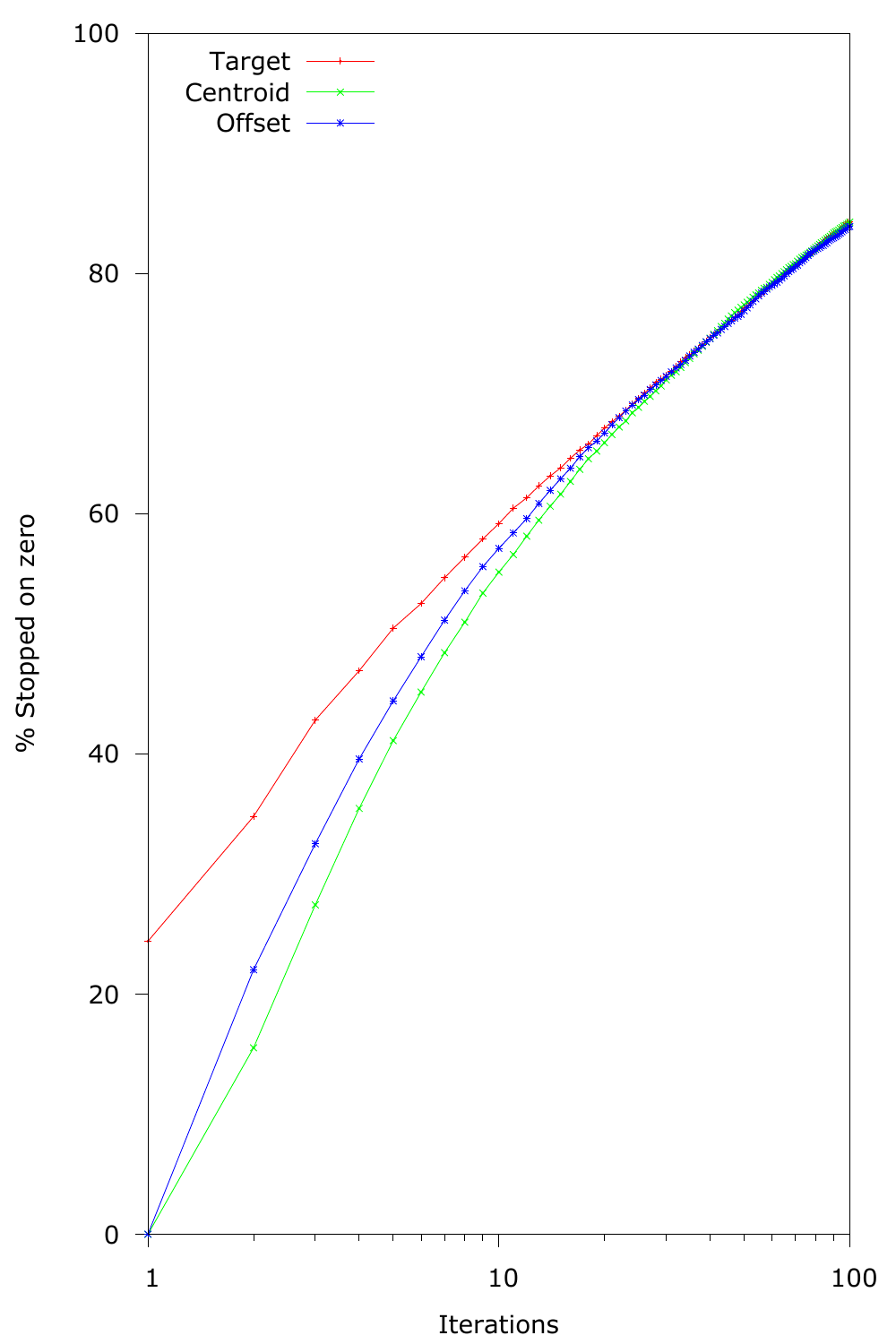}
	\label{fig:zeroexperiment}
	}
	\quad
	\subfigure[Mean relative improvement between iteration $i-1$ and $i$ on a log scale.]{
		\includegraphics[width=55mm]{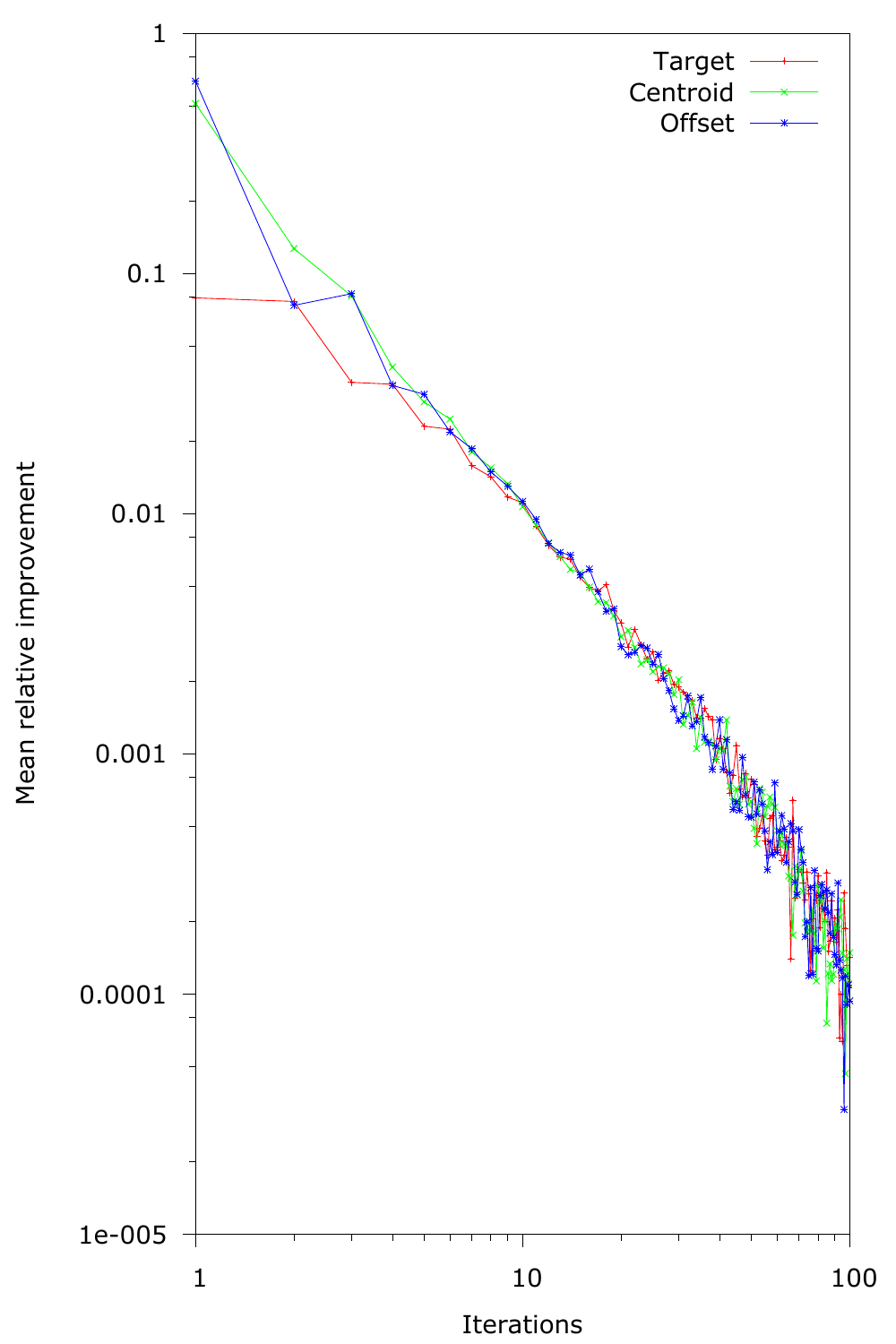}
	\label{fig:improveexperiment}
	}
	\caption{Effect of increasing the number of iterations on the algorithm performance.}
\end{figure}

Because a large portion of samples could not be improved when $\omega_0$ is set to the target vector, we want to examine how general this behavior is. For our second experiment we varied the number of iterations from $1$ to $100$ for $q=0.6$ and the three different choices of $\omega_0$. We drew $10,000$ samples from the $8$D Simplex and counted how often a sample $t$ ended in a Banzhaf vector with a zero power player for increasing number of iterations. This is a cumulative figure: when the algorithm stops on sample $t$ at $i$ iterations performing $i+1$ iterations will also stop the algorithm. Figure \ref{fig:zeroexperiment} contains the results, with the number of iterations set on a log scale. Additionally, figure \ref{fig:improveexperiment} shows the mean relative improvement between iterations for this experiment, with the improvement also set on a log scale.

Figure \ref{fig:zeroexperiment} shows that the choice of $\omega_0$ only has a significant effect on the operation of the algorithm for the first $20$ iterations. At $5$ iterations the difference between starting at the target and starting in the centroid of the ordered simplex is still $10\%$, but this drops to less than $1\%$ when the number of iterations is more than $25$. We can also see that starting in the centroid of the ordered Simplex actually has the best chance of avoiding a zero power player initially. On the other hand, setting $\omega_0$ as target gives a $20\%$ chance to start with a zero power player, which results in an unimprovable first guess.

In general, it is obvious that the number of cases where the algorithm stops due to a $0$ in the Banzhaf vector grows with the number of iterations, going from more than $40\%$ at $5$ iterations to above $80\%$ for $100$ iterations. Further, the effect of an iteration is only significant for the first $4$ iterations. This means that generally the algorithm will not manage to converge to the target value, and it also suggests a possible area of improvement.

\section{Improvements }
In order to factor out the premature stop due to zeros in the Banzhaf vector we propose and evaluate three possible changes:
\begin{enumerate}
	\item Restarting the algorithm on different weights. By applying a different transformation to the weights when a player has $0$ power the algorithm can `restart' on a new guess.
	\item Imposing a minimum coalition size. By enforcing a minimum coalition size powerless players in the original algorithm can become powerful by helping a coalition reach its size quota.
	\item Introducing a scaling factor in the coefficient computation. By enforcing a strictly positive value for powerless players in the ratio computation we can continue with the algorithm even when a player has no power.
\end{enumerate}
The following sections describe these three improvements in more detail.

\subsection{Multiple Start}
As we have seen in figure \ref{fig:zeroexperiment}, the algorithm \ref{alg:base} quickly encounters the situation where computing the ratio results in a divide by zero from a player with $0$ power in its Banzhaf vector. To continue with the algorithm at this point, we may try to find a different transformation to perform on the weights. The resulting new weights are in effect a second guess, leading to a restart with better initial weights. This procedure can then be repeated as often as we can make better guesses, resulting in the Multiple Start version of the algorithm.

The restart guess should be a transformation on the weights discovered during past iterations to make it meaningful. If the restart guess was not based off earlier results, it would be possible to simply apply the original algorithm to this better guess instead. It is also not useful to restart on a set of weights that has already been tried, since the algorithm is deterministic.

A promising place to start from is the set of weights that produced the best power distribution thus far. We propose to use the Offset Target procedure on this set of best weights to obtain a new set of weights halfway between the centroid of the regular simplex and the best weights, as shown in algorithm \ref{alg:restart}. Our motivation for this choice is threefold:
\begin{itemize}
	\item We know from the exploration of the Banzhaf vector locations that the majority is located near the center, and thus approaching our target from the center out gives the best chance of finding a closer vector to jump to.
	\item In our initial experiments, the Offset Target heuristic provided good final results, with the best worst case error and a robust average case. 
	\item We can reason about its positive effect for some non-improvable samples. For example, consider the case when the first player has all the power because he has a weight $\frac{1}{2}<w_1<1-\frac{1}{n}$. Then this operation brings his weight below $\frac{1}{2}$. Because the first player had the most weight, this means that all weights are reduced to below $\frac{1}{2}$ and no player can be a dictator in the transformed sample.
\end{itemize}

We expect this version of the algorithm to never return a solution that is worse than the standard version for the same number of iterations, since it uses the standard version up until the first restart. Restarting may also provide a bigger benefit when the standard algorithm frequently cannot improve beyond the initial guess, which is for values of $q$ near $0$, $0.5$ or $1$. We know that in the limit all inputs tend to go to a power vector with a zero power player, but if the standard algorithm can perform many iterations, there may not be any room for further improvement by starting somewhere else. Thus we expect this algorithm is likely to produce the biggest improvement early on.

\begin{algorithm}
	\caption{LW with Restarts ($v$, $\text{target}$, $\omega_0$, $\text{maxIterations}$, $\text{maxDistance}$)}
	\label{alg:restart}
	\begin{algorithmic}[1]
	\Require {Target is a normalized vector of size $n$, with $n > 0$.}
	\State $\text{weight} \gets \omega_0$
	\State $\text{bestweight} \gets \omega_0$
	\State $\text{iterations} \gets 0$
	\State $\text{distance} \gets \infty$
	\Repeat
		\State $\text{banzhafIndex} \gets \text{calculateBanzhaf}(v, \text{weights})$
		\State For each player $i$: $\displaystyle \text{ratio}(i) \gets \frac{\text{banzhafIndex}(i)}{\text{target}(i)}$
		\If{$\text{containsZero}(\text{banzhafIndex})$}
			\State For each player $i$: $\displaystyle \text{weight}(i) \gets \frac{\text{bestweight}(i) + \frac{1}{n}}{2}$
		\Else
			\State For each player $i$: $\displaystyle \text{weight}(i) \gets \frac{\text{weight}(i)}{\text{ratio}(i)}$
		\EndIf
		\State $\text{distance} \gets \text{distance}(\text{banzhaf}, \text{target})$
		\If{$\text{distance was improved}$}
			\State $\text{bestweight} \gets \text{weight}$
		\EndIf
		\State $\text{iterations} \gets \text{iterations} + 1$
	\Until{$\text{distance} < \text{maxDistance} \lor \text{iterations} > \text{maxIterations}$}
	\Ensure {weight is a vector of size n}
	\end{algorithmic}
\end{algorithm}

\subsection{Minimum Coalition Size}
Instead of recovering from a situation where a player has no power, we may try to manipulate the algorithm so that it will not produce weights where a player has no power. Laruelle and Widgr\'{e}n evaluate the EU for different rules, and some of these rules have valuation functions that impose a minimum coalition size (for example, QM+ requires $10$ out of $15$ countries to vote in favor). One aspect of the minimum coalition size is that it gives power to players that normally would not have any.

To show that this is the case, assume a weight distribution where only $w_1>0$. Even though player $1$ has all the power, if we impose minimum coalition sizes of $2$ and up, player 1 must form coalitions with others to win. In this way, the other players get some power. Note that for minimum coalition sizes smaller than $n$ the others still get less power than player $1$, since only he is present in all of these coalitions. More generally, the number of possibly winning coalitions is $2^n-1$ without restrictions, or $\sum\limits_{i=m}^{n} \binom{n}{i}-1$ for an imposed minimum coalition size $m$.

As a consequence, while this method can resolve cases where the Banzhaf vector would otherwise contain zeros, it also restricts the number of Banzhaf vectors available. For example, the vector $[\frac{3}{5}\:\frac{1}{5}\:\frac{1}{5}]$ has four winning coalitions $[1 0 0], [0 1 1], [1 1 0], [1 0 1]$. But the first coalition has size $1$, so it would not be winning if minimum coalition size were set to $2$.

Therefore, we expect that imposing minimum coalition sizes sometimes produces worse results than the standard algorithm. Furthermore, when the minimum coalition size increases, fewer Banzhaf vectors are available, resulting in larger errors. We can thus expect the ideal minimum coalition size to be relatively small.

\subsection{Scaling Factor}
Another way to work around the case where a player has no power is to always use a strictly positive update coefficient $r$. We may introduce a scaling factor $s$ in both the numerator and denominator of the calculation, i.e. $r_i=\frac{\beta_i+s}{\beta_\text{fair}+s}$. This ensures that for $\beta_i=0$ the coefficient is still a positive real.

The secondary effect of this scaling factor is that the magnitude of adjustment is smaller, which leads to slower convergence. This may on its own also lead to better results, since it reduces the risk of overshooting the target vector. On the other hand, it will take more iterations to get to the desired value. Another risk is that the algorithm no longer manages to jump out of local minima.

Since the scaling factor changes the operation of the algorithm, the effect of this change is somewhat uncertain. We may expect better results when the standard algorithm stops in one iteration due to a zero power player. Its effect in general is more difficult to predict. We may need more iterations to get the same or better results as the standard version, if convergence is indeed slower. In other words, there is a possibility of returning worse results in general.

\section{Evaluation}
\begin{figure}
	\centering
	\subfigure[Coalition]{
		\includegraphics[width=36mm]{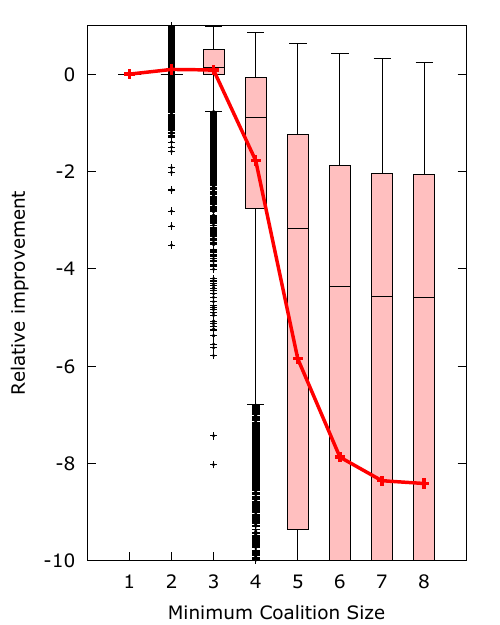}
	}
	\subfigure[Restarts]{
		\includegraphics[width=36mm]{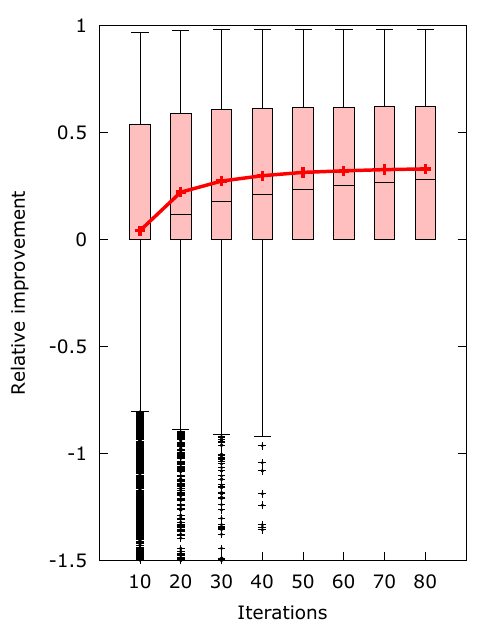}
	}
	\subfigure[Scaling]{
		\includegraphics[width=36mm]{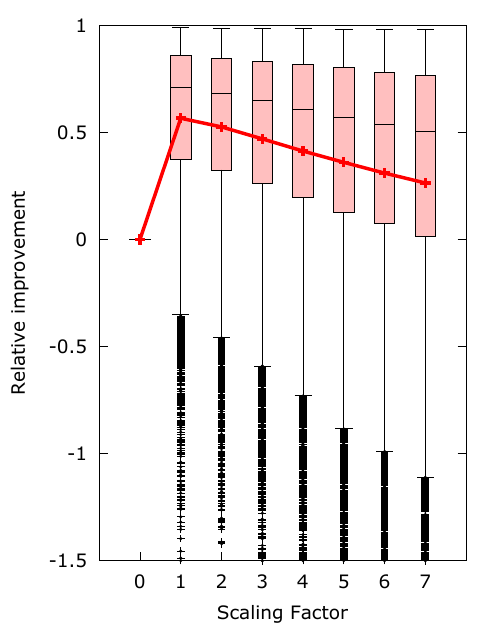}
	}
	\caption{Relative improvement for different parameter settings on the suggested improvements. In all cases $q=0.5$, $\omega_0=t$. Results for $10,000$ samples.}
	\label{fig:parametereval}
\end{figure}

Since each suggested improvement has its own parameters, we should look at the effect of these parameters in isolation. To examine what relative improvement can be obtained, we drew $10,000$ samples from the ordered $8$D Simplex and compared the result with that of the standard algorithm, with parameters $q=0.5$, $\omega_0$ set to target, and run for $50$ iterations. For the restart improvement we increased the number of iterations performed from $10$ through $80$. For the minimum coalition improvement we varied the minimum coalition size from $1$ through $8$. And finally, for the scaling improvement we varied the scaling factor from $0$ through $7$. Other parameters set equal to the standard version. The results can be seen in figure \ref{fig:parametereval}.

The results show that indeed not all improvements are universal improvements. Both minimum coalition and scaling have samples that have worse distance than the base algorithm. The restart algorithm also needs at least the same amount of iterations to obtain a universal improvement.

Minimum coalition has the largest fluctuation in performance, going from significant improvement at $2$ and $3$ to almost always producing worse results at $5$ and higher. This confirms our prediction that the imposed minimum coalition size should be kept small.

Restarting shows very significant improvements (more than $0.2$ in the mean) when the number of iterations is $20$ or higher. This means that the restart version can typically be used with fewer iterations than the base version of the algorithm, which can be important if runtime is a consideration.

The size of the scaling factor does seem to have a large impact on the quality of improvement, but even for the factor set to $7$ the algorithm shows very significant improvement. Improvement for $1$ and $2$ exceeds $0.5$. It may be that the high factors obtain the same quality results when the algorithm is given more iterations, but in general it is better to use a small scaling factor.

\begin{figure}[ht]
	\begin{center}
	\subfigure[Minimum coalition size $3$, scaling factor $0$, no restarts]{
		\includegraphics[width=34mm]{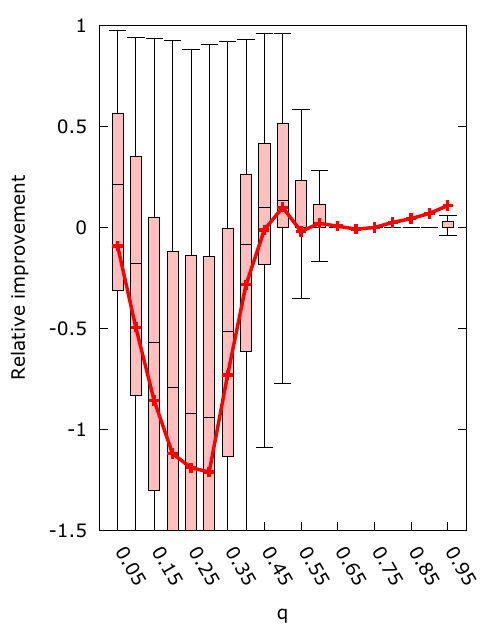}
	}
	\quad
	\subfigure[Minimum coalition size $1$, scaling factor $0$, restarts]{
		\includegraphics[width=34mm]{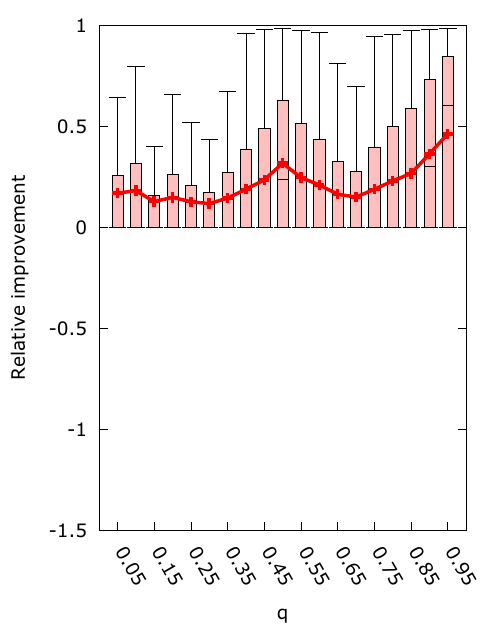}
	}
	\quad
	\subfigure[Minimum coalition size $1$, scaling factor $0.4$, no restarts]{
		\includegraphics[width=34mm]{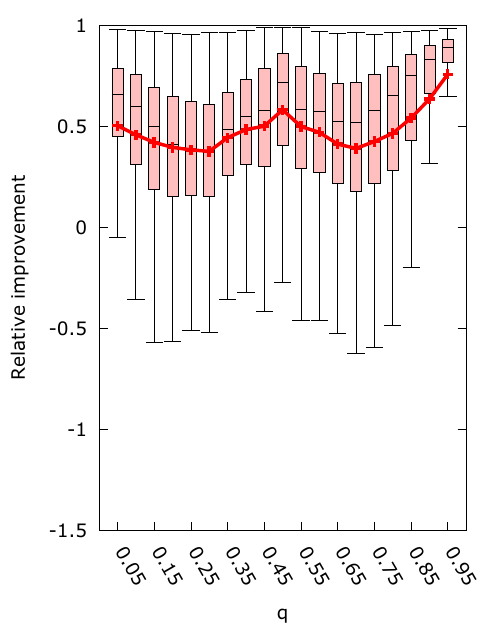}
	}
	\end{center}
	\caption{Effect of $q$ on distance to target for the suggested improvements compared with the original implementation, $\omega_0=t$}
	\label{fig:imprexperiments}
\end{figure}

A second experiment to look at the improvement compared to the input parameter $q$ was performed to determine how the new algorithms compare to the original for sensible parameter settings. We again varied $q$ from $0.05$ to $0.95$ and applied all the algorithms on $5,000$ new $8$D samples. Each algorithm was allowed $50$ iterations, minimum coalition size was set to $3$ and the scaling factor was set to $0.4$ for the respective algorithms. Figure \ref{fig:imprexperiments} presents the results.

From figure \ref{fig:imprexperiments} it is immediately clear that a scaling factor is the superior improvement, giving the best mean improvement for the entire range of $q$. We can also see that the general asymmetry of the algorithm with respect to the choice of $q$ is maintained, with lower $q$ behaving worse than higher $q$ for all algorithms. Interestingly, all algorithms show a peak in improvement when $q=0.5$, supporting our intuition that it is the ideal selection of $q$. For scaling, setting $q=0.5$ actually produced the best absolute results. However, all algorithms actually make the best improvement for very high $q$. This does not have much practical use, since the original algorithm performs very poorly for such high $q$.

We see further confirmation of the effect of $q$ on average coalition sizes when looking at the results of minimum coalition. There, results are strongly negative for low $q$, while almost no samples are affected for high $q$. This can be explained by considering that for these low $q$ the average coalition size is smaller than the minimum imposed coalition size, while for high $q$ almost no set of weights produces a game with coalitions of size $3$.

In order to be able to say something about the improvement of the proposed additions with respect to the error we performed an experiment with tuned parameters for each. Every algorithm was run on $5,000$ samples of $8$D Simplex, with parameters set to produce the best results: $50$ iterations, $q=0.6$ except for scaling which performed better for $q=0.5$, $\omega_0$ set to centroid, minimum coalition size $3$, scaling factor $0.4$. Table \ref{tab:finalimprovements} contains the results. The first column shows the algorithm under consideration. The second column details how many of the samples were improved compared to the base, while the third contains the amount of samples that were made worse. Column four shows how often an addition produced strictly the best result compared to all the others, including the base version. The fifth column presents the average distance to the target vector. Column six displays the average distance to the best known power vector. The last row of the table shows what the results would be like if we could always pick the best algorithm for a sample.

\begin{table}
	\centering
	\begin{tabular}{ l | r r r r r }
		Algorithm & \% Improved & \hspace{5mm}\% Worse & \hspace{5mm}\% Best & \hspace{2mm} Error upper & \hspace{2mm} Error lower \\
		\hline
		Base	  &		  -  &		 -  &		-  & $0.0801$ &	$0.0520$ \\
		Coalition & $23.6\%$ & $18.6\%$ &  $3.6\%$ & $0.0761$ & $0.0481$ \\
		Restart	  & $32.9\%$ &	$0.0\%$ &  $3.0\%$ & $0.0643$ & $0.0363$ \\
		Scaling   & $86.2\%$ & $12.0\%$	& $80.0\%$ & $0.0357$ & $0.0078$ \\
		\hline
		Sum/Best  &       -  &       -  & $86.5\%$ & $0.0344$ &	$0.0065$ \\
	\end{tabular}
	\vspace{3mm}
	\caption{Effect of the improvements for ideal parameter settings.}
	\label{tab:finalimprovements}
\end{table}

In column four we can see that in $86.5\%$ of the samples, one of the improvements produced a game that was closest to $t$. For most samples scaling produced the closest result, however all three improvements have samples they performed best on. The base algorithm never produced a game closest to $t$ since restarting was given the same number of iterations as the base version, which means it always produced at least the same output. However we can see in column three that both coalition and scaling do produce output that is worse than the base version in more than $10\%$ of the samples.

The table shows that all additions reduce the error on average, however the magnitude of improvement can only be considered significant for scaling. Additionally, for scaling the upper bound on the error is below $\frac{7}{185}=0.0378\ldots$, and the lower bound is below $0.5\%$ of the maximum $d_1$, which is $1.75$ for $8$ players. Therefore we can say that the error made by scaling is not significant in size. Compared to the base algorithm with ideal parameters, introducing a scaling factor improves the lower bound performance by a factor $6.5$.

The other approaches have their own strengths, as restarting does not make the result worse, and minimum coalition actually returns the best result more often than restarting does. In a sense the approaches can be seen to complement each other. If we always take the best result, the improvement compared with scaling is still another $20\%$ in the lower bound which indicates it could be worthwhile to find a new version of the algorithm that combines the effect of the three approaches in some way.

\section{Conclusion and Future Work}
The algorithm by Laruelle and Widgr\'en works quite well in most cases, but it has some major shortcomings. Our proposals remove the possibility of the algorithm getting stuck in a case where one or more zeroes are in the weight vector, and experiments show that our scaling factor algorithm also improve the average approximation performance. However, in some cases it performs worse than the original algorithm. Our multiple start proposal performs at least as well as the original algorithm, but improves the solution not nearly as much as the scaling factor algorithm. Further work could be done to find an algorithm that improves on these proposals: either by giving a better worst-case performance or by improving the average case approximation, or both.

Our experiments also show that the algorithm is not anytime: an iteration often improves the solution, but it could also deteriorate. Our improvements do not counter that, other than storing the best found solution. This is also something that could be researched in the future.

\bibliographystyle{plain}
\bibliography{paper}

\end{document}